\documentclass[aps,prb,twocolumn,showpacs,amsmath,amssymb,superscriptaddress]{revtex4-1}

\usepackage{graphicx}
\usepackage{subfigure}
\usepackage{epsfig}
\usepackage{dcolumn}
\usepackage{bm}
\usepackage{ulem}
\usepackage{color}

\def\be{\begin{equation}}       \def\ee{\end{equation}}
\def\bea{\begin{eqnarray}}      \def\eea{\end{eqnarray}}
\def\ba{\begin{array} }
\def\ea{\end{array} }
\def\bnum{\begin{enumerate} }
\def\enum{\end{enumerate}}

\def\nn{\nonumber}
\def\pa{\partial}
\def\=>{\Rightarrow}
\def\>{\rightarrow}
\def\A{\uparrow}
\def\V{\downarrow}

\def\eye2{Fathbb{I}}

\def\Eq#1{Eq.~(\ref{#1})}
\def\Fig#1{Fig.~\ref{#1}}

\renewcommand{\>}{\rangle}

\begin{document}

\title{Topological p+ip superconductivity in doped graphene-like single-sheet materials BC$_3$}
\author{Xi Chen}
\affiliation{Department of Physics and State Key Laboratory of Low Dimensional Quantum Physics, Tsinghua University, Beijing, 100084, China}
\author{Yugui Yao}
\affiliation{School of Physics, Beijing Institute of Technology, Beijing, 100081, China}
\author{Hong Yao}
\email{yaohong@tsinghua.edu.cn}
\affiliation{Institute for Advanced Study, Tsinghua University, Beijing, 100084, China}
\affiliation{Collaborative Innovation Center of Quantum Matter, Beijing, China}
\author{Fan Yang}
\email{yangfan_blg@bit.edu.cn}
\affiliation{School of Physics, Beijing Institute of Technology, Beijing, 100081, China}
\author{Jun Ni}
\email{junni@mail.tsinghua.edu.cn}
\affiliation{Department of Physics and State Key Laboratory of Low Dimensional Quantum Physics, Tsinghua University, Beijing, 100084, China}
\affiliation{Collaborative Innovation Center of Quantum Matter, Beijing, China}

\begin{abstract}
We theoretically study exotic superconducting phases in graphenelike single-sheet material BC$_3$ doped to its type-II Van Hove singularity whose saddle point momenta are {\it not} time-reversal-invariant. From combined renormalization group analysis and RPA calculations, we show that the dominant superconducting instability induced by weak repulsive interactions is in the time-reversal-invariant $p+ip$ pairing channel because of the interplay among dominant ferromagnetic fluctuations, subleading spin fluctuations at finite momentum, and spin-orbit coupling. Such time-reversal-invariant $p+ip$ superconductivity has nontrivial $\mathbb{Z}_2$ topological invariant. Our results show that doped BC$_3$ provides a promising route to realize a genuine 2D helical $p+ip$ superconductor.
\end{abstract}


\maketitle
\section{\label{intro}Introduction}
Many exciting discoveries of topological quantum states of matter[\onlinecite{wenbook}] have been made in recent years, including topological insulators protected by time-reversal symmetry[\onlinecite{qi-zhang-11,hasan-kane-10,qi-hughes-zhang-08}], quantum
Hall effect in graphene [\onlinecite{yuanbo-05},\onlinecite{novoselov-05}], and quantum anomalous Hall effect [\onlinecite{qkxue-13}]. Nonetheless, a class of
intrinsic superconductors with fully-gapped bulk excitations but robust gapless boundary excitations dubbed as ``topological superconductors'' [\onlinecite{kitaev-09,schnyder-08,Qi2}] have not been unambiguously identified in nature even though enormous efforts have been devoted
into their discoveries. Among them, the two-dimensional (2D) $p+ip$ superconductors are of special importance, including both chiral and helical $p+ip$ superconductors.  The intrinsic chiral $p+ip$ superconductivity is believed to exist in Sr$_2$RuO$_4$ [\onlinecite{Sigrist-95,Mackenzie,sri-kapitulnik-kivelson,ronny-13}]. The magnetic vortices of chiral $p+ip$ superconductor support Majorana zero modes [\onlinecite{Tewari-Sarma-Lee-08},\onlinecite{Read-Green}], which obey non-Abelian statistics [\onlinecite{Read-Green},\onlinecite{Ivanov}] and which are believed to be a promising tool for topological quantum computation [\onlinecite{kitaev-03},\onlinecite{Nayak}]. However,
evidence for Sr$_2$RuO$_4$ being a fully gapped chiral $p+ip$ superconductor is still not definitive [\onlinecite{review-12}]. As a close cousin of chiral $p+ip$ superconductor, the helical $p+ip$ superconductor is time-reversal-invariant, and supports helical Majorana modes along its boundary [\onlinecite{Qi2}].

Graphene [\onlinecite{novoselov-04},\onlinecite{castroneto-09}] has attracted special attention as candidate materials harboring unconventional superconductivity (SC) induced by electronic interactions [\onlinecite{uchoa-07},\onlinecite{sri-kivelson-scalapino}] when doped away from half-filling. In particular, at about 1/4 electron or hole doping, the Fermi level is at the Van Hove singularity (VHS) where density of states (DOS) is logarithmically divergent and where it was proposed in previous theoretical analysis that SC with $d+id$ pairing and relatively high transition temperature may be induced by repulsive interactions [\onlinecite{chubukov-12,dhlee-121,ronny-12}]. Nevertheless, the arguably more interesting triplet $p+ip$ pairing was not reported there. It was pointed out recently by two of us [\onlinecite{yao-13}] that the absence of $p+ip$ triplet pairing in graphene at the VHS is mainly due to the fact that its saddle point momenta ${\bf K}$ are time-reversal invariant (TRI), namely ${\bf K}=-{\bf K}$. Such Van Hove saddle points are called ``type-I''. For systems at type-I VHS, triplet pairing potential at saddle points must vanish due to the Pauli exclusion principle; consequently, triplet pairing is normally suppressed. The concept of type-II VHS was introduced in Ref. [\onlinecite{yao-13}]; for type-II VHS, Van Hove saddle point momenta ${\bf K}$ are not TRI, namely ${\bf K}\neq -{\bf K}$. It was shown that systems with type-II VHS are promising arenas to look for topological $p+ip$ triplet pairings[\onlinecite{yao-13}].

In this paper, we propose the BC$_3$ doped to its type-II VHS as a highly promising material to look for $p+ip$ topological superconductivity (TSC). We construct an effective model of BC$_3$ and perform renormalization group (RG) analysis to investigate competing orders in BC$_3$ at type-II VHS. We show that the dominant instability is SC when considering weak repulsive interactions between electrons. Due to the type-II VHS and the resulting strong ferromagnetic fluctuation in about 1/8 doped BC$_3$, the triplet pairings are more favored than singlet. The interplay between ferromagnetic fluctuation and spin fluctuations at finite momenta yields $p+ip$ SC. More interestingly, for BC$_3$ with spin-orbit coupling (SOC), we show that the helical $p+ip$ superconductor with nontrivial $\mathbb{Z}_2$ topological invariant is the leading instability. This result obtained by RG analysis is consistent with the one from calculations within random phase approximation (RPA). We believe the graphenelike BC$_3$ doped to its type-II VHS could provide a promising arena to realize genuinely two-dimensional helical $p+ip$ SC.

\section{\label{results}Methods and Results}

\begin{figure}
\begin{center}
\includegraphics[width=8.2cm]{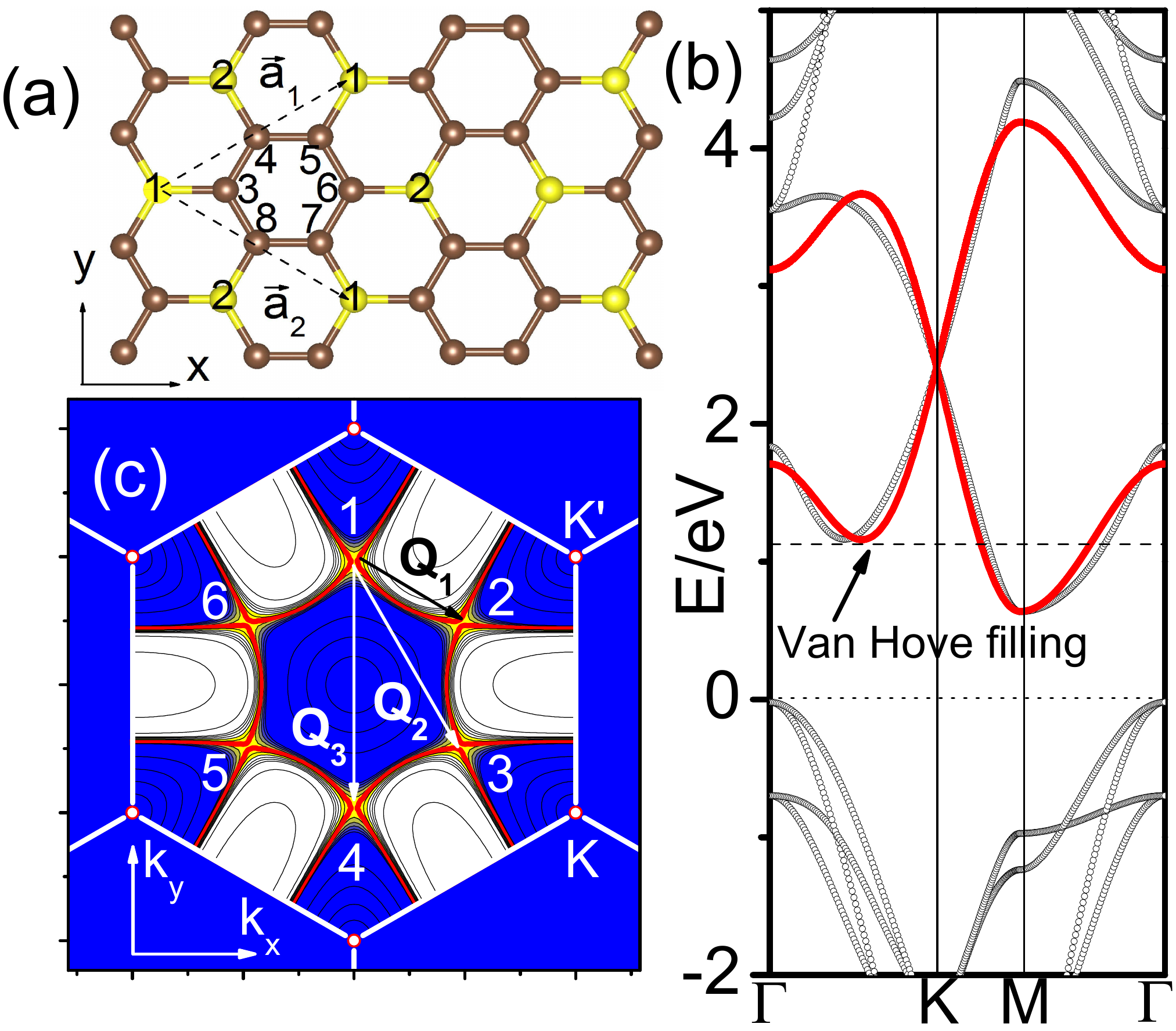}
\end{center}
\caption{ (a) The atomic structures of BC$_3$. The yellow (light) circles denote the boron atoms while the brown (dark) circles denote the carbon atoms. (b) The band structures of BC$_3$. The open circles represent the density functional results. The red (dark) lines are the tight-binding band structures calculated with $t_1=0.62$eV and $t_3=-0.38$eV. (c) The FS of doped BC$_3$. The white areas denote the occupied states while the blue (dark) areas represent the unoccupied states. The saddle points highlighted in yellow (light) are connected by the vectors $\vec{Q}_1$, $\vec{Q}_2$ and $\vec{Q}_3$.  }
\label{fig1}
\end{figure}

\subsection{Material}
BC$_3$ is a graphenelike genuine 2D material as shown in \Fig{fig1}(a), which was successfully fabricated in experiments [\onlinecite{Yanagisawa}]. We compute its band structure via density functional theory [\onlinecite{Kresse1},\onlinecite{Kresse2}] and the results are shown in \Fig{fig1}(b). Undoped BC$_3$ is a band insulator with band gap $\Delta_g\sim 0.5$eV. With slight electron doping, the Fermi level moves into the first conduction band which mainly consists of the $p_z$ orbital of boron atoms [\onlinecite{Miyamoto}]. It was shown that doping may be achieved through chemical absorption with lithium adatoms [\onlinecite{Chen}]. We use the number of electrons doped per site, namely $x$, to quantify the doping concentration. With small $x$, the Fermi surface (FS) consists three electrons pockets around M points. At the critical doping $x_c \sim 1/8$, its FS goes through a Lifshitz transition at which it has six saddle points inside the Brillouin zone, as shown in \Fig{fig1}(c). This is exactly type-II VHS[\onlinecite{yao-13}]. To the best of our knowledge, BC$_3$ is the first genuinely-2D material with hexagonal symmetry which realizes type-II VHS. Previous first-principle calculations indicate that the FS close to the type-II VHS results in strong magnetic fluctuations[\onlinecite{Chen}]. In the limit of weak interactions, it is known that SC is the leading instability. As a close interplay between magnetism and SC is expected, we investigate the phase diagram of the system at type-II VHS as a function of interactions by addressing the following issues:
(i) What kind of magnetic ordering, if any, occurs in the doped BC$_3$ when the repulsive interactions are relatively strong? (ii) What pairing symmetry is in the superconducting phases of doped BC$_3$ in the limit of weak interactions? (iii) What are deep connections between the pairing symmetry and nature of magnetic ordering in this system?

Because of the type-II VHS in the FS, ferromagnetic fluctuations are expected to strong. As a consequence, we expect that ferromagnetic magnetic ordering develops when short-range repulsive interactions are relatively strong. When the interaction strength is below a critical value, the quantum fluctuations spoil the long-range magnetic order and unconventional SC should emerge. When the long-range magnetic orders are absent, the magnetic fluctuations are strong in this system and peak at $\vec{Q}=\vec{0}$ and $\vec{Q}_i (i=1\sim3)$. Here $\vec{Q}_i$ are visualized in \Fig{fig1}(c). These fluctuations could mediate attractive interaction between quasiparticles and lead to unconventional SC [\onlinecite{Scalapino89}]. The pairing symmetry of resultant SC depend on a subtle interplay between magnetic fluctuations and on the crystal symmetry. According to the point-group symmetry of BC$_3$, SC can occur in $s$, $p_x$, $p_y$, $d_{x^2-y^2}$, $d_{xy}$, and $f$ channels. Here $p_x$ and $p_y$ form a 2D irreducible representation of the hexagonal system; consequently they have degenerate pairing instability. The same is true for $d_{x^2-y^2}$ and $d_{xy}$ pairings. For such degenerate pairing channels, we can show that the $p+ip$ pairing ($d+id$ pairing) always has lower energy than nodal $p$-wave pairing ($d$-wave pairing) because the FS can be fully gapped by it (See Supplementary Material). There are two types of triplet $p+ip$ pairings, namely chiral and helical $p+ip$ pairings, which have degenerate energy when SOC is absent. Nonetheless, there is only one type of singlet $d+id$ pairing which is always chiral. In summary, the possible pairing channels are $s$, chiral $p+ip$, helical $p+ip$, chiral $d+id$, and $f$. Among them, the chiral/helical $p+ip$ and $f$ channels are triplet which are mediated by the spin fluctuation at $\vec{Q}=0$, namely ferromagnetic fluctuations. The $d+id$ channel is derived from the spin fluctuations at $\vec{Q}_i(i=1\sim3)$. The competitions between different spin fluctuations play a decisive role in the competitions between different pairing symmetries. Due to the type-II VHS in doped BC$_3$, the triplet channels are not suppressed. Consequently, we can not neglect the ferromagnetic fluctuation as in Ref. [\onlinecite{chubukov-12}]; otherwise we will mistake the $d+id$ channel as the leading instability. As we will show in the following, the ferromagnetic fluctuation and spin fluctuation at $\vec{Q}_1$ mutually lead to the $p+ip$ pairings in doped BC$_3$.

\subsection{Model}
We consider the following Hubbard model to describe the main physics of the first conduction band:
\bea
H = \sum_{ij} (t_{ij}c_{i\sigma}^{\dag}c_{j\sigma}+H.c.)+\sum_i Uc^\dag_{i\A}c_{i\A}c^\dag_{i\V}c_{i\V},
\eea
where $c_{i\sigma}$ annihilates a $p_z$ electron of borons with spin polarization $\sigma=\A,\V$ on site $i$ and $U$ is the Hubbard interaction which mimics the short-range Coulomb repulsions. Here $t_{ij}=t_1,t_2,t_3$ label
effective electron hoppings between nearest, next-nearest, and third neighbor boron atoms, which are mediated by carbon atoms and their values are obtained by fitting the band structures obtained from {ab inito} calculations. Here are optimized parameters: $t_1=0.62$eV, $t_3=-0.38$eV, and $t_2$ is about one order of magnitude smaller than $t_1$ and $t_3$. A negligible $t_2$ is due to destructive interference between two hopping loops, {\it e.g.} along two loops ``3-4-5'' and ``3-8-7-6-5'', as shown in \Fig{fig1}(a). Hereafter we neglect $t_2$ for simplicity. The tight-binding band structures with optimized hopping parameters are shown in Fig. 1(b), which fits quite well with the one computed from the first principle calculations. From first principles calculations, we estimate that $U\sim0.7$eV.

\subsection{RG analysis}
In the limit of weak interactions, low energy physics are dominated by electrons near the FS. Moreover, in a 2D system at VHS, electrons near those saddle points dominate the logarithmically divergent DOS. Consequently, we consider only the fermions in the patchs around the saddle points. With this approximation, the low energy behavior of the system can be described by the effective action in continuum limit:
\bea\label{equa_1}
S&=&\int d\tau d^2r \sum_{\alpha=1}^6 \sum_\sigma\psi_{\alpha\sigma}^{\dag}[ \pa_\tau-\varepsilon_{\alpha}(i\pa_x,i\pa_y) - \mu]\psi_{\alpha \sigma} \nn\\
&&~~~~~~~~+\sum_{\alpha \beta \gamma} \sum_{\sigma \sigma^{\prime}} \frac{1}{2}g_{\alpha \beta \gamma\delta}\psi_{\alpha \sigma}^{\dag}\psi_{\beta \sigma^{\prime}}^{\dag}\psi_{\gamma \sigma^{\prime}}\psi_{\delta\sigma},
\eea
where $\psi^\dag_{\alpha\sigma}$ creates an electron in patch $\alpha=1,\cdots,6$ with spin polarization $\sigma$, $\varepsilon_{\alpha}$ represents electronic dispersions in patch $\alpha$, and $\mu=0$ describes the system exactly at the Van Hove filling. Note that $\delta$ above is implicitly determined from $\alpha\beta\gamma$ by momentum conservation. Note that we neglect the SOC here because it is expected to be weak in material consisting of such
light atoms as borons and carbons; but we shall consider it when we dope heavy metallic atoms into the system.

\begin{widetext}
The $g_{\alpha \beta \gamma\delta}$ in \Eq{equa_1} describes various interactions allowed by the symmetries of the system under consideration.  Because of the lattice symmetries of the hexagonal system BC$_3$, there are totally only nine inequivalent interaction parameters: $g_{1}=g_{1441}$, $g_{2}=g_{1436}$, $g_{3}=g_{1425}$, $g_{4}=g_{1414}$, $g_{5}=g_{1313}$,
$g_{6}=g_{1331}$, $g_{7}=g_{1212}$, $g_{8}=g_{1221}$, $g_{9}=g_{1111}$.
To investigate possible phase transitions as temperature decreases, we study how
interactions flow using RG equations derived from gradually integrating out electrons between a decreasing $\omega$ and the ultraviolet cutoff $\Lambda$[\onlinecite{RMP}]. We introduce dimension-less interaction parameters $g_{\alpha\beta\gamma\delta}\to \nu_0 g_{\alpha\beta\gamma\delta}$ and derive the one-loop RG flow equation[\onlinecite{Furukawa}] for this hexagonal system at the type-II Van Hove singularity as follows:
\begin{small}
\begin{equation}\label{equa3}
\begin{aligned}
&\frac{dg_1}{dy}=-d_\textrm{pp}^{14}(g_1^2+2g_1^2+2g_3^2+g_4^2)+d_\textrm{ph}^{14}g_1^2+2d_\textrm{ph}^{11}(g_4g_9+2g_6g_7+2g_5g_8-g_1g_4-4g_6g_8), \\
&\frac{dg_2}{dy}=-2d_\textrm{pp}^{14}(g_1g_2+g_2g_3+g_3g_4)+2d_\textrm{ph}^{13}g_2g_6+2d_\textrm{ph}^{12}(g_2g_8+g_3g_7-2g_2g_7), \\
&\frac{dg_3}{dy}=-d_\textrm{pp}^{14}(2g_1g_3+2g_2g_4+g_2^2+g_3^2)+2d_\textrm{ph}^{12}g_3g_8+2d_\textrm{ph}^{13}(g_3g_6+g_2g_5-2g_3g_6), \\
&\frac{dg_4}{dy}=-2d_\textrm{pp}^{14}(g_1g_4+2g_2g_3)+2d_\textrm{ph}^{11}(g_4g_9+2g_5g_7)+2d_\textrm{ph}^{14}(g_4g_1-g_4^2), \\
&\frac{dg_5}{dy}=-2d_\textrm{pp}^{13}g_5g_6+d_\textrm{ph}^{11}(2g_5g_9+2g_4g_7+g_7^2+g_5^2)+2d_\textrm{ph}^{13}(g_5g_6+g_2g_3-g_5^2-g_3^2), \\
&\frac{dg_6}{dy}=-d_\textrm{pp}^{13}(g_5^2+g_6^2)+d_\textrm{ph}^{13}(g_2^2+g_6^2)+2d_\textrm{ph}^{11}(g_7g_8+g_5g_9+g_4g_8+g_5g_6+g_1g_7-g_8^2-g_6^2-g_6g_9-2g_1g_8), \\
&\frac{dg_7}{dy}=-2d_\textrm{pp}^{12}g_7g_8+2d_\textrm{ph}^{11}(g_7g_9+g_5g_7+g_4g_5)+2d_\textrm{ph}^{12}(g_7g_8+g_2g_3-g_7^2-g_2^2), \\
&\frac{dg_8}{dy}=-d_\textrm{pp}^{12}(g_7^2+g_8^2)+2d_\textrm{ph}^{12}(g_3^2+g_8^2)+2d_\textrm{ph}^{11}(g_1g_5+g_5g_8+g_4g_6+g_6g_7+g_7g_9-2g_1g_6-2g_6g_8-g_8g_9), \\
&\frac{dg_9}{dy}=-d_\textrm{pp}^{11}g_9^2+d_\textrm{ph}^{11}(g_4^2+2g_5^2+2g_7^2+g_9^2)+2d_\textrm{ph}^{11}(g_1g_4+2g_5g_6+2g_7g_8-g_1^2-2g_6^2-2g_8^2), \\
\end{aligned}
\end{equation}
\end{small}
where $y\equiv \log^2(\Lambda/\omega)$ is the flow parameter.
\end{widetext}

The above $d$-functions $d^{\alpha\beta}_\textrm{pp}=d_{pp}(\vec{P}_{\alpha}+\vec{P}_{\beta})$ and $d^{\alpha\beta}_\textrm{ph}=
d_{ph}(\vec{P}_{\alpha}-\vec{P}_{\gamma})$ are defined as $d^{\alpha\beta}_\textrm{pp}=\frac{2}{\nu_0}\frac{\pa\chi^{\alpha\beta}_\textrm{pp}}{\pa y}$ and $d^{\alpha\gamma}_\textrm{ph}=\frac{2}{\nu_0}\frac{\pa\chi^{\alpha\gamma}_\textrm{ph}}{\pa y}$, in which $\chi_\textrm{pp}^{\alpha \beta}\equiv\chi_\textrm{pp}(\vec{P}_{\alpha}+\vec{P}_{\beta},\omega)$ and
$\chi_\textrm{ph}^{\alpha\gamma}\equiv \chi_\textrm{ph}(\vec{P}_{\alpha}-\vec{P}_{\gamma},\omega)$ are the susceptibilities of noninteracting electrons in the electron-electron and electron-hole channels, respectively. Since these functions depend implicitly on $\varepsilon_{\alpha}$(the electronic dispersions in the patch $\alpha$), we can expand $\varepsilon_{\alpha}$ as $\varepsilon_{\alpha}(\delta k_x,\delta k_y)=\delta k_i \delta k_j/(2m_{ij}) +O(\delta k^3)$  with $\delta\vec{k}=\vec{k}-\vec{P}_{\alpha}$ [$\vec{P}_{\alpha}$ denotes the saddle point momentum in patch $\alpha$, as shown in Fig.1(c)]. We label eigenvalues of the mass matrix $m_{ij}$ as $m_1$ and $m_2$ ($m_1\approx 1.6$eV and $m_2\approx 1.2$eV for BC$_3$ at type-II VHS). The DOS per patch $\rho(\omega)$ diverges logarithimcally: $\rho(\omega)\approx \nu_0 \log(\Lambda/\omega)$, where $\Lambda$ is order of band width, and $\omega$ is the energy away from VHS. $\nu_0$ is a numerical factor, which depends on the band structure through $\nu_0=\sqrt{m_1m_2}/(4\pi^2)$.

For $\omega\ll\Lambda$, these non-interacting susceptibilities are given by
\bea
\chi_\textrm{pp}(\vec Q_3)\approx a_3 \frac{\nu_0}{2}\log^2(\Lambda/\omega), ~ &\ \chi_\textrm{ph}(\vec 0) \approx \nu_0\log(\Lambda/\omega),~~~~~ \nn\\
\chi_\textrm{pp}(\vec Q_2)\approx \bar a \nu_0\log(\Lambda/\omega), ~ &\ \chi_\textrm{ph}(\vec Q_1)\approx a \nu_0\log(\Lambda/\omega), \nn\\
\chi_\textrm{pp}(\vec Q_1)\approx \bar a \nu_0\log(\Lambda/\omega), ~ &\ \chi_\textrm{ph}(\vec Q_2)\approx a \nu_0\log(\Lambda/\omega), \nn\\
\chi_\textrm{pp}(\vec 0)\approx\frac{\nu_0}2\log^2(\Lambda/\omega), ~&\ \chi_\textrm{ph}(\vec Q_3)\approx a_3 \nu_0\log(\Lambda/\omega)\nn,
\eea
where $\vec Q_i\equiv \vec P_{i+1}-\vec P_1$, $\bar a$ and $a$ are functions of the mass ratio $\kappa=m_1/m_2$, while $0<a_3<1$ depends on the details of dispersions around Van Hove saddle points. Note that $\chi_\textrm{ph}(\vec Q_1)$ and $\chi_\textrm{ph}(\vec Q_2)$  have identical leading logarithmical divergent behaviour, which is required by the lattice symmetry of the hexagonal system we consider. Similarly, $\chi_\textrm{ph}(\vec Q_1)$ and $\chi_\textrm{ph}(\vec Q_2)$ have identical leading logarithmical divergent behaviour.

The detailed behavior of $d_{pp}$ and $d_{ph}$ depends on specifics of the band structure. Nonetheless, their have the asymptotic forms: as $y\to 0$, $d\to 1$
for all channels; as $y\to \infty$, $d_\textrm{ph}(\vec 0)\to1/\sqrt{y}$, $d_\textrm{ph}(\vec Q_{1/2})\to a/\sqrt{y}$, $d_\textrm{ph}(\vec Q_3)\to a_3/\sqrt{y}$, $d_\textrm{pp}(\vec Q_3)\to a_3$ and $d_\textrm{pp}(\vec Q_{1/2})\to\bar a/\sqrt{y}$. Following Refs. [\onlinecite{Furukawa,chubukov-12,yao-13}], we model these $d$
functions using the following analytic forms: $d_\textrm{ph}(\vec 0)\approx \frac1{\sqrt{y+1}}$, $d_\textrm{ph}(\vec Q_3)\approx \frac{a_3}{\sqrt{y+a_3^2}}$, $d_\textrm{pp}(\vec Q_3)\approx\frac{1+ a_3 y}{1+y}$ and $d_\textrm{pp}(\vec Q_{1/2})\approx \frac{\bar a}{\sqrt{y+\bar a^2}}$ and , all of
which fulfill their asymptotic behaviors. The leading instability does not sensitively depend on the values of $\bar a$ and $a_3$, which are order of one. Hereafter we assume $\bar a=a_3=1.0$ for simplicity. $d_\textrm{ph}(\vec Q_1)$ and $d_\textrm{ph}(\vec Q_2)$ deserve special attention since the paring symmetry of superconductivity may sensitively depend on which one of them has larger sub-logarithmic values even though they have identical leading logarithmical behavior. So we employ an approximate analytic forms for them:  $d_\textrm{ph}(\vec Q_i)\approx\frac{a}{\sqrt{y+b_i{y}^{1/2}+a^2}}$ ($i=1,2$), where $b_i$ is introduce to describe the sub-logarithmic behavior of $d_\textrm{ph}(\vec Q_i)$ which determine the competition between $p$ and $f$-wave superconductors. If we set $b_1=b_2=0$, namely model them similarly as other $d$-functions, we would obtain $g_{1425}(y)=g_{1436}(y)$ for all $y$ even though there is no symmetry which dictates this property. Moreover, if $g_{1425}(y)=g_{1436}(y)$ for all $y$, $p+ip$ and $f$ pairing are always degenerate, which is also not required by any symmetry of the system. Hereafter we consider finite but different $b_1$ and $b_2$.

The interactions $g_1,\cdots,g_9$ defined above have the asymptotic behavior $g_i\sim\frac{G_i}{y_c-y}$ as $y\rightarrow y_c$. The susceptibility exponents for various types of broken symmetries can be expressed as the linear combination of $G_i$. The
susceptibility exponents for $s$-wave paring, $p$-wave paring, $d$-wave paring, $f$-wave paring, spin density wave (SDW) with momentum ${\vec Q}_1$ (denoted as SDW1), SDW with ${vec Q}_2$ (SDW2), ferromagnetism, charge density wave (CDW) with ${\vec Q}_1$ (denoted as CDW1), and CDW with ${\vec Q}_2$ (CDW2) are given as: 
\begin{equation}\label{equa3}
\begin{aligned}
\gamma_s&\ =-2(G_1+2G_2+2G_3+G_4) \\
\gamma_p&\ =-2(G_1+G_2-G_3-G_4) \\
\gamma_d&\ =-2(G_1-G_2-G_3+G_4) \\
\gamma_f&\ =-2(G_1-2G_2+2G_3-G_4) \\
\gamma_\textrm{SDW1}&\ =2(G_3+G_8)\cdot d_\textrm{ph}(\vec Q_1,y_c) \\
\gamma_\textrm{SDW2}&\ =2(G_2+G_6)\cdot d_\textrm{ph}(\vec Q_2,y_c) \\
\gamma_\textrm{FM}&\ =2(G_4+G_5+G_7+G_9)\cdot d_\textrm{ph}(\vec 0,y_c) \\
\gamma_\textrm{CDW1}&\ =2(G_3+G_8-2G_2-2G_7)\cdot d_\textrm{ph}(\vec Q_1,y_c) \\
\gamma_\textrm{CDW2}&\ =2(G_2+G_6-2G_3-2G_5)\cdot d_\textrm{ph}(\vec Q_2,y_c)
\end{aligned}
\end{equation}

From the RG flow equations, we obtain the low energy effective interaction $g_{\alpha \beta \gamma\delta}(y)$ and calculate susceptibilities of various broken symmetries of the interacting electrons. These susceptibilities show asymptotic form $\chi\sim (y_c-y)^{-\gamma}$ when $y\rightarrow y_c$. A positive exponent $\gamma$ leads to divergent susceptibility when $y\rightarrow y_c$, which indicates the ordering tendency with decreasing temperature. The most positive $\gamma$ tell us the leading instability.

\begin{figure}
\begin{center}
\includegraphics[width=8.2cm]{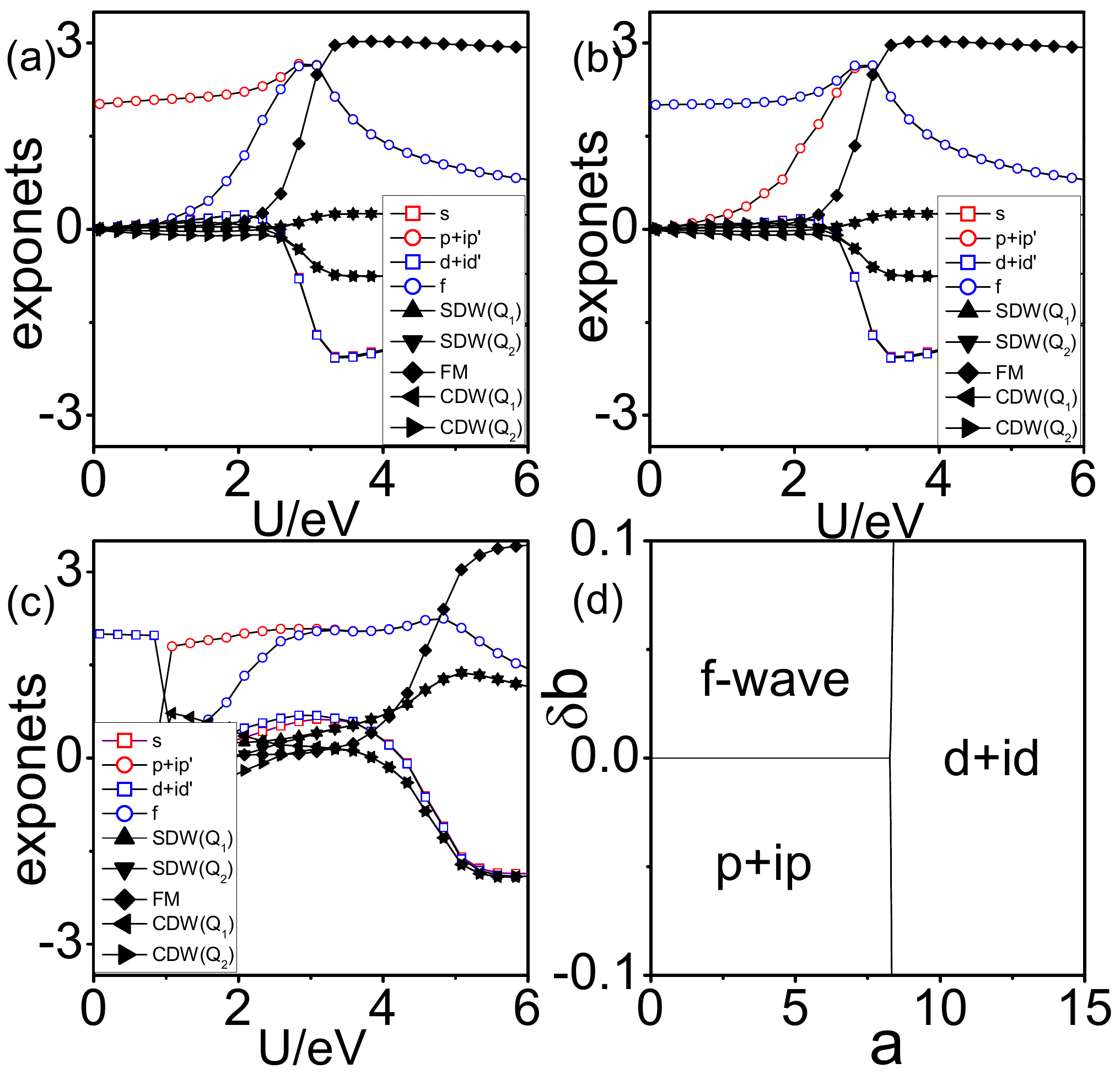}
\end{center}
\caption{ (a)$\sim$(c): The evolution of susceptibility exponents of various types of broken symmetries as a function of Hubbard $U$ for three different choice of parameters: (i) $a=2.0$, $b_1=0.025$, and $b_2=0.075$; (ii) $a=2.0$, $b_1=0.075$, and $b_2=0.025$; and (iii) $a=9.0$, $b_1=0.025$ and $b_2=0.075$; (d): The phase diagrams as a function of $a$ and $\delta b\equiv b_1-b_2$ with $b_1+b_2=0.1$. The results are calculated for $U=0.7$eV.   }
\label{fig2}
\end{figure}

The interplays between magnetic fluctuations are modeled with $d_\textrm{ph}(\vec Q_1)$ and $d_\textrm{ph}(\vec Q_2)$ in the flow equation. We take an approximate analytic forms for them: $d_\textrm{ph}(\vec Q_i)\approx\frac{a}{\sqrt{y+b_i{y}^{1/2}+a^2}}$ ($i=1,2$) [\onlinecite{Furukawa},\onlinecite{yao-13}]. The parameter $a$ describes the extent of FS nesting and it determines the competition between ferromagnetic spin fluctuation and spin fluctuations at finite momenta. When the FS is perfectly nested, we have $a\to \infty$. As a result, the spin fluctuations at $\vec{Q}_1$ and $\vec{Q}_2$ are maximized at $d_\textrm{ph}(\vec Q_i)\to 1$. In such cases, singlet pairing is favored. If, on the contrary, the FS nesting is weak, we would have small $a$ and thus reduced spin fluctuations at $\vec{Q}_1$ and $\vec{Q}_2$, which favors triplet pairing. The $b_i$ ($i=1,2$) describe competition between spin fluctuations at $\vec{Q}_1$ and $\vec{Q}_2$, since smaller $b_i$ leads to stronger fluctuation $d_\textrm{ph}(\vec Q_i)$. As we will show in the following, this competition determines the competition between $p+ip$ and $f$-wave pairing. Note that the spin fluctuation at $\vec{Q}_3$ is not that essential in the sense that it dose not affect the main physics we consider.

We proceed by addressing the instability of the FS in doped BC$_3$ as a function of Hubbard $U$. We obtain various susceptibility exponents as a function of Hubbard $U$ for three cases with different $a$ or $b_i$ parameters, as shown in \Fig{fig2}(a-c). The ferromagnetism is the leading instability when the Hubbard $U$ is beyond a critical value, which is approximately 3eV in \Fig{fig2}(a-b) and 5eV in \Fig{fig2}(c). For weaker $U$, the flow of interaction parameters always favors superconducting phases, as expected and shown in \Fig{fig2}(a-c). For doped BC$_3$, it was estimated that $U\sim0.7$eV, which indicates the leading instability is SC.

The pairing symmetry depends on the relative strengths among magnetic fluctuations at momentum $\vec{0}$ (namely ferromagnetic spin fluctuation), at $\vec{Q}_1$, and at $\vec{Q}_2$. To be relevant to BC$_3$, we set the interaction strength $U=0.7eV$
and obtain the phase diagram as a function of $a$ and $\delta b\equiv b_1-b_2$, as shown in \Fig{fig2}(d). This phase diagram suggests that when $a\leq8$ ($a>8$), the spin fluctuations at $\vec{Q}_1$ and $\vec{Q}_2$ are weaker (stronger) than the ferromagnetic fluctuation, which leads to triplet (singlet) pairing. While the leading pairing symmetry of the singlet pairing in the phase diagram is always in the $d+id$ channel,  that for the triplet pairings can be either $p+ip$ or $f$, which is determined by negative or positive $\delta b$ respectively. For the tight-binding model of BC$_{3}$, our calculations yield $a\approx 1.3<8$ and $\delta b<0$, which leads to $p+ip$ pairing in the system when it is doped to the type-II VHS.

\subsection{RPA result}
\begin{figure}[tbp]
\includegraphics[width=8.2cm]{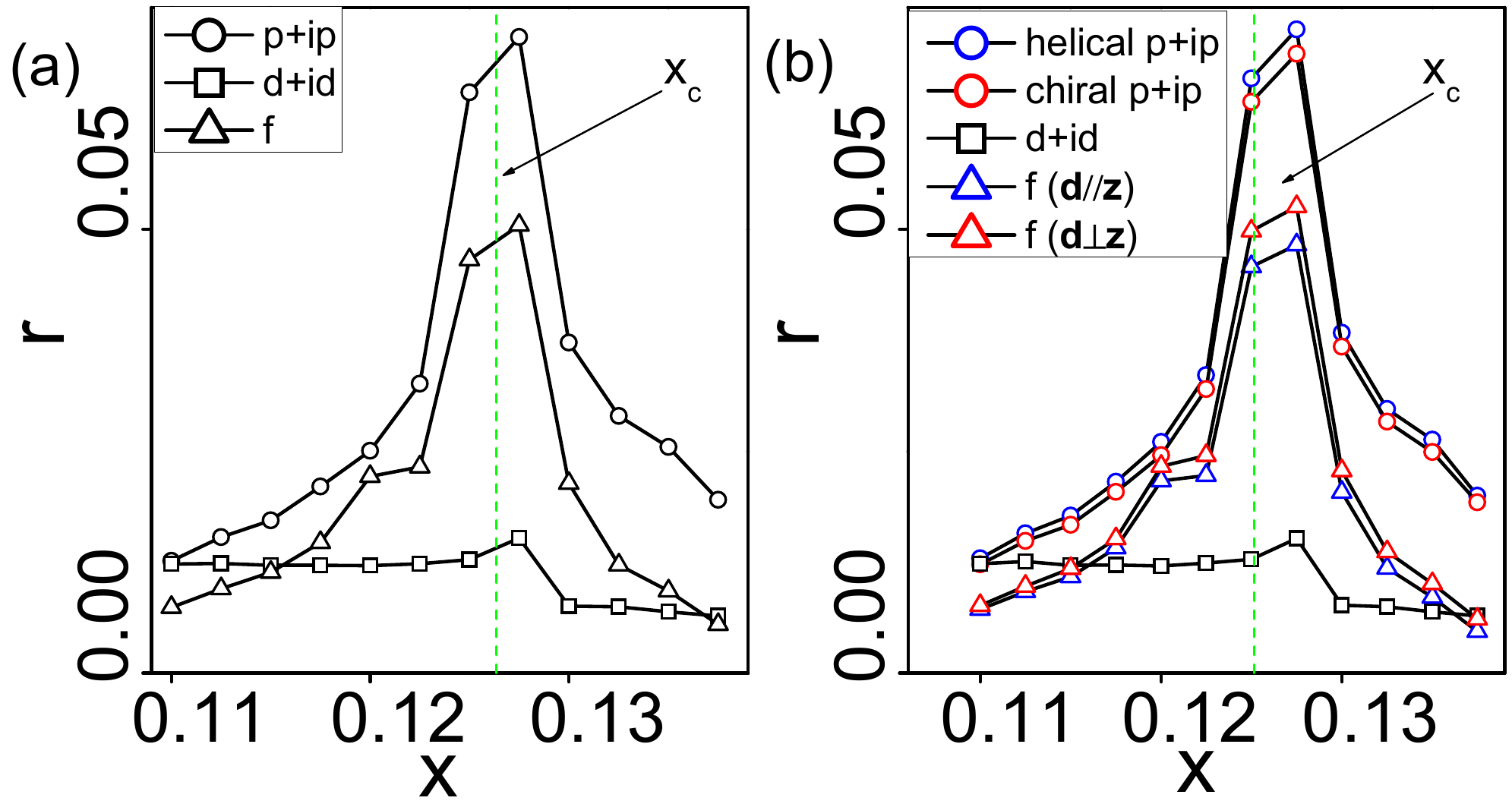}
\caption{(Color online) Doping-dependence of the largest eigenvalues of the linearized gaps equations near $T_c$ for different pairing symmetries without (a) and with (b) Kane-Mele SOC ($\lambda=0.05t_{1}$).}
\label{Fig:doping}
\end{figure}

The RG analysis above shows that the leading instability in the BC$_3$ doped exactly to its type-II VHS at $x_c\approx 0.127$ is the $p+ip$ triplet pairing. To investigate its broken symmetry phases away from the VHS, we have also performed a RPA based study for the pairing symmetries of the system near $x_c$ for supplement. Standard multi-orbital RPA approach [\onlinecite{Scalapino89},\onlinecite{liu-2013}] is adopted in our study for the doping regime $x\in(0.11,0.135)$ with $U=0.8t_1$. Via exchanging spin fluctuations, electrons near the FS acquire an effective pairing interaction $V_\textrm{eff}$, from which one obtains the linearized gap equation which has solutions in various pairing channels. The leading instability occurs in the pairing channel with the largest eigenvalue $r$ of the linearized gap equation with $T_c\sim t_1e^{-1/r}$. The doping dependence of $r$ in various pairing channels near $x_c$ is shown in \Fig{Fig:doping}(a), which suggests that the odd parity $p+ip$ and $f$-wave pairings are the leading and subleading pairing symmetries respectively. This result is consistent with the RG analysis above.

\subsection{Helical TSC}
The leading triplet $p+ip$ pairing obtained in BC$_3$ is characterized by its $\vec d_{\vec k}$-vector defined through $\langle\psi^\dag_{\vec k s}\psi^\dag_{-\vec k s'}\rangle\propto (\vec d_{\vec k}\cdot \vec \sigma \sigma^y)_{ss'}$. Without SOC, the $p+ip$ pairings with different $\vec d_{\vec k}$-vectors are exactly degenerate, which includes both time-reversal breaking chiral TSC and time-reversal-invariant helical TSC[\onlinecite{Zhang}], as shown in the appendix. A finite SOC can lift the degeneracy between the helical and chiral $p+ip$ pairings. In BC$_3$, the inversion symmetry allows us to consider the Kane-Mele SOC, whose strength is parameterized by $\lambda$, in the tight-binding Hamiltonian and then perform the RPA calculations to obtain the leading pairing symmetry for weak $U$ and $\lambda$. From RPA calculations, triplet $p+ip$ pairing is the leading instability when $\lambda=0$. For weak but finite $\lambda$, while both the $\left(p+ip\right)_{\left(\uparrow\downarrow+\downarrow\uparrow\right)}$ chiral TSC (with $\vec d_{\vec k}\parallel \mathbf{z}$ ) and the $\left(p+ip\right)_{\left(\uparrow\uparrow\right)};\left(p-ip\right)_{\left(\downarrow\downarrow\right)}$ helical TSC (with $\vec d_{\vec k}\perp \mathbf{z}$ ) are possible, our RPA results select the latter as the leading pairing symmetry, as shown in \Fig{Fig:doping}(b) with a weak $\lambda=0.05t_1$.

\section{Conclusions and Discussions}
In summary, we have performed combined RG analysis and RPA calculations for doped BC$_3$ at type-II Van Hove singularity and show that the strong ferromagnetic fluctuation mediates triplet pairings (either $p+ip$ or $f$-wave pairing) for weak repulsive interactions. The competition between $p+ip$ and $f$-wave SC depends on the competition between spin fluctuations at $\vec Q_1$ and $\vec Q_2$. The relatively stronger spin fluctuation at $\vec Q_1$ favor $p+ip$ pairing as the leading instability with a relatively-high transition temperature enhanced by the VHS. A weak Kane-Mele type SOC favors helical $p+ip$ pairing over the chiral one. The gap structure of the $p+ip$ pairing can be detected by the phase-sensitive dc-SQUID devices[\onlinecite{SQUID}]. Further more, such helical $p+ip$ SC respects time-reversal-symmetry with the hallmark that it supports helical gapless Majorana edge modes which are robust against disorder as long as the time-reversal-symmetry is preserved and which should be detectable in STM measurements. Besides being a promising material to look for a genuine 2D helical $p+ip$ SC with nontrivial $\mathbb{Z}_2$ topological invariant [\onlinecite{dhlee-122,Nagaosa,shoucheng-14,yang-14}], doped BC$_3$ might have potential applications in areas such as topological quantum computations as well as realizing emergent supersymmetry[\onlinecite{Grover,sslee,yao-14}] in the future.

{\it Acknowledgements}: This work was supported in part by the NSFC under Grant Nos. 11374175 and 11174171 (X.C. and J.N.), and by the National Thousand-Young-Talent Program of China and the NSFC under Grant No. 11474175 (H.Y.). Y.Y. was supported by the MOST Project of China (Grants No. 2014CB920903) and the NSFC (Grant Nos. 11174337 and 11225418). F.Y. was supported by the NSFC (Grant Nos. 11274041 and 11334012) and the NCET program under the Grant No. NCET-12-0038.

\appendix

\section{The Ginzburge-Landau free energy of $p+ip$ superconductors}

To derive the Landau-Ginzburg free energy, we start with the partition function
$Z=\int D[\bar{\psi}\psi]\exp(-\int \mathcal{L}(\bar{\psi}\psi))$ where

\begin{eqnarray}\label{equa1}
\mathcal{L}
&=&\mathcal{L}_0+\mathcal{L}_{int},\\
&=&\sum_{\alpha=1}^6 \sum_{\sigma} \bar{\psi}_{\alpha \sigma}[\partial_{\tau} + \varepsilon_{\alpha}(\delta \vec{k}) - \mu]\psi_{\alpha \sigma} \nn\\
&&~~~~~~+\frac{1}{2}\sum_{\alpha \beta} \sum_{\sigma \sigma^{\prime}} g_{\alpha \tilde{\alpha} \tilde{\beta} \beta}\bar{\psi}_{\alpha \sigma}\bar{\psi}_{\tilde{\alpha} \sigma^{\prime}}\psi_{\tilde{\beta} \sigma^{\prime}}\psi_{\beta\sigma},
\end{eqnarray}
where $\tilde{\alpha}$ means $\vec{P}_{\tilde{{\alpha}}}=-\vec{P}_{\alpha}$. Notice
that $g_{1}=g_{1441}$, $g_{2}=g_{1436}$, $g_{3}=g_{1425}$, $g_{4}=g_{1414}$,
we can write $\mathcal{L}_{int}$ in the matrix form as:
\begin{equation}\label{equa1}
\mathcal{L}_{int}=\frac{1}{2}\begin{bmatrix}
                                \bar{\psi}_{1 \sigma}\bar{\psi}_{4 \sigma^{\prime}} \\
                                \bar{\psi}_{2 \sigma}\bar{\psi}_{5 \sigma^{\prime}} \\
                                \bar{\psi}_{3 \sigma}\bar{\psi}_{6 \sigma^{\prime}} \\
                                \bar{\psi}_{4 \sigma}\bar{\psi}_{1 \sigma^{\prime}} \\
                                \bar{\psi}_{5 \sigma}\bar{\psi}_{2 \sigma^{\prime}} \\
                                \bar{\psi}_{6 \sigma}\bar{\psi}_{3 \sigma^{\prime}} \\
                            \end{bmatrix}^{T}\begin{bmatrix}
                                                g_1 & g_2 & g_3 & g_4 & g_3 & g_2 \\
                                                g_2 & g_1 & g_2 & g_3 & g_4 & g_3 \\
                                                g_3 & g_2 & g_1 & g_2 & g_3 & g_4 \\
                                                g_4 & g_3 & g_2 & g_1 & g_2 & g_3 \\
                                                g_3 & g_4 & g_3 & g_2 & g_1 & g_2 \\
                                                g_2 & g_3 & g_4 & g_3 & g_2 & g_1 \\
                                            \end{bmatrix}
                            \begin{bmatrix}
                                \psi_{4 \sigma^{\prime}}\psi_{1 \sigma} \\
                                \psi_{5 \sigma^{\prime}}\psi_{2 \sigma} \\
                                \psi_{6 \sigma^{\prime}}\psi_{3 \sigma} \\
                                \psi_{1 \sigma^{\prime}}\psi_{4 \sigma} \\
                                \psi_{2 \sigma^{\prime}}\psi_{5 \sigma} \\
                                \psi_{3 \sigma^{\prime}}\psi_{6 \sigma} \\
                            \end{bmatrix}.
\end{equation}
The eigenvectors of the interaction matrix above are:
\begin{equation}\label{equa1}
\begin{aligned}
\Delta_{s}
&\ =\frac{\Delta}{\sqrt{6}}\begin{pmatrix}
                                    1, & 1, & 1, & 1, & 1, & 1 \\
                                  \end{pmatrix}, \\
\Delta_{p_x}
&\ =\frac{\Delta}{\sqrt{4}}\begin{pmatrix}
                                    0, & 1, & 1, & 0, & -1, & -1 \\
                                  \end{pmatrix}, \\
\Delta_{p_y}
&\ =\frac{\Delta}{\sqrt{12}}\begin{pmatrix}
                                    2, & 1, & -1, & -2, & -1, & 1 \\
                                  \end{pmatrix}, \\
\Delta_{d_{x^2-y^2}}
&\ =\frac{\Delta}{\sqrt{4}}\begin{pmatrix}
                                    0, & 1, & -1, & 0, & 1, & -1 \\
                                  \end{pmatrix}, \\
\Delta_{d_{xy}}
&\ =\frac{\Delta}{\sqrt{12}}\begin{pmatrix}
                                    2, & -1, & -1, & 2, & -1, & -1 \\
                                  \end{pmatrix}, \\
\Delta_{f}
&\ =\frac{\Delta}{\sqrt{6}}\begin{pmatrix}
                                    1, & -1, & 1, & -1, & 1, & -1 \\
                                  \end{pmatrix},
\end{aligned}
\end{equation}
which represent $s$, $p_x$, $p_y$, $d_{x^2-y^2}$, $d_{xy}$, and $f$-wave paring symmetries, respectively. The corresponding eigenvalues are given by
\begin{equation}\label{equa1}
\begin{aligned}
&\lambda_s=g_1+2g_2+2g_3+g_4, \\
&\lambda_p=g_1+g_2-g_3-g_4, \\
&\lambda_d=g_1-g_2-g_3+g_4, \\
&\lambda_f=g_1-2g_2+2g_3-g_4,
\end{aligned}
\end{equation}
which describe the paring strength between the electrons.

We focus on the $p$-wave superconductors which are proved to be the leading
instability in the main text. To decouple the quartic interactions, we introduce the order parameter in the patch space:
\begin{equation}\label{equa1}
\Delta
=[i(\vec{\Delta}_1\cdot\vec{\sigma})\sigma_y]\otimes[P_x]+[i(\vec{\Delta}_2\cdot\vec{\sigma})\sigma_y]\otimes[P_y]
\end{equation}
in which $\vec{\Delta}_i$ ($i=1,2$) are complex vectors and $\vec{\sigma}$ are Pauli matrix.
The matrix $P_x$ and $P_y$ are:
\begin{equation}\label{equa1}
\begin{aligned}
P_x&=\frac{1}{2}\mathrm{diag}\begin{pmatrix}
                        0, &  1, &  1 \\
                      \end{pmatrix}, \\
P_y&=\frac{1}{12}\mathrm{diag}\begin{pmatrix}
                        2, &  1, &  -1 \\
                      \end{pmatrix}.
\end{aligned}
\end{equation}
Using the Hubbard-Stratonovich transformation, we obtain:
\begin{equation}\label{equa1}
\mathcal{L^{\prime}}
=\bar{\Psi}\left(
             \begin{array}{cc}
               G_{+}^{-1} & \Delta \\
               \Delta^{\dagger} & G_{+}^{-1} \\
             \end{array}
           \right)\Psi+\frac{|\vec{\Delta}_{1}|^2+|\vec{\Delta}_{2}|^2}{|\lambda_p|},
\end{equation}
where $\Psi=(\psi_{1\uparrow}~\psi_{1\downarrow}~\psi_{2\uparrow}~\psi_{2\downarrow}~ \psi_{3\uparrow}~\psi_{3\downarrow}~\psi_{4\uparrow}^{\dagger}~\psi_{4\downarrow}^{\dagger}~ \psi_{5\uparrow}^{\dagger}~\psi_{5\downarrow}^{\dagger}~\psi_{6\uparrow}^{\dagger}~ \psi_{6\downarrow}^{\dagger})^T$. Here, $G_{+}$ and $G_{-}$ are particle and hole propagators with the form
$G_{\pm}^{-1}=-i\omega_n\pm[\varepsilon(\delta \vec{k})-\mu]$. They
are diagonal in the patch space. By integrating out the fermion operators,
we get the effective action:
\begin{equation}\label{equa1}
\mathcal{L^{\prime\prime}}
=-\mathrm{Tr}\ln\left(
             \begin{array}{cc}
               G_{+}^{-1} & \Delta \\
               \Delta^{\dagger} & G_{-}^{-1} \\
             \end{array}
           \right)+\frac{|\vec{\Delta}_{1}|^2+|\vec{\Delta}_{2}|^2}{|\lambda_p|}.
\end{equation}

\begin{widetext}
From expanding the first term in $\mathcal{L^{\prime\prime}}$ to the quartic term in $\Delta$, we get:
\begin{eqnarray}\label{equa1}
\mathrm{Tr}\ln\left(
             \begin{array}{cc}
               G_{+}^{-1} & \Delta \\
               \Delta^{\dagger} & G_{-}^{-1} \\
             \end{array}
           \right)
&\approx&-\mathrm{Tr}[G_{+}\Delta G_{-}\Delta^{\dagger}]
-\frac{1}{2}\mathrm{Tr}[G_{+}\Delta G_{-}\Delta^{\dagger}G_{+}\Delta G_{-}\Delta^{\dagger}]\nn\\
&=&-\mathrm{Tr}[G_{+} G_{-}]\mathrm{Tr}[\Delta\Delta^{\dagger}]-\frac{1}{2}\mathrm{Tr}[G_{+}G_{-}G_{+}G_{-}]\mathrm{Tr}[\Delta\Delta^{\dagger}\Delta \Delta^{\dagger}],\nn
\end{eqnarray}
where the trace means integration over $\vec k$. Due to the rotational
symmetry, $\mathrm{Tr}[G_{+} G_{-}]$ is identical for all the patches and
could be factored out of the trace over the patch space. Using the identity
$\mathrm{Tr}[P_x^2]$=1/2, $\mathrm{Tr}[P_y^2]$=1/2, $\mathrm{Tr}[P_xP_y]$=0,
$\mathrm{Tr}[P_x^4]$=1/8, $\mathrm{Tr}[P_y^4]$=1/8, $\mathrm{Tr}[P_x^2P_y^2]$=$\mathrm{Tr}[P_xP_yP_xP_y]$=1/24 and $[i(\vec{\Delta}_1\cdot\vec{\sigma})\sigma_y][i(\vec{\Delta}_1\cdot\vec{\sigma})\sigma_y]^{\dagger}$
=$|\vec{\Delta}_1|^2\mathrm{I}+i(\vec{\Delta}_1\times\vec{\Delta}_1^*)\cdot\vec{\sigma}$
, we obtain 
\begin{eqnarray}\label{equa1}
\mathrm{Tr}[\Delta\Delta^{\dagger}]
&=&\frac{1}{2}\mathrm{Tr}[|\vec{\Delta}_1|^2\mathrm{I}+i(\vec{\Delta}_1\times\vec{\Delta}_1^*)\cdot\vec{\sigma}]
+\frac{1}{2}\mathrm{Tr}[|\vec{\Delta}_2|^2\mathrm{I}+i(\vec{\Delta}_2\times\vec{\Delta}_2^*)\cdot\vec{\sigma}]
=|\vec{\Delta}_1|^2+|\vec{\Delta}_2|^2,\\
\mathrm{Tr}[\Delta\Delta^{\dagger}\Delta \Delta^{\dagger}]
&=&\frac{1}{8}\{|\vec{\Delta}_1|^4-(\vec{\Delta}_1\times\vec{\Delta}_1^*)^2+|\vec{\Delta}_2|^4-(\vec{\Delta}_2\times\vec{\Delta}_2^*)^2\} +\frac{1}{6}\{|\vec{\Delta}_1|^2|\vec{\Delta}_2|^2-(\vec{\Delta}_1\times\vec{\Delta}_1^*)\cdot(\vec{\Delta}_2\times\vec{\Delta}_2^*)\}\\
\nn\\
&&~~~~+\frac{1}{24}\{(\vec{\Delta}_1\cdot\vec{\Delta}_2^*)^2-(\vec{\Delta}_1\times\vec{\Delta}_2^*)\cdot(\vec{\Delta}_1\times\vec{\Delta}_2^*)+h.c.\}.
\end{eqnarray}
Since $\vec{\Delta}_1\times\vec{\Delta}_1^*=-\vec{\Delta}_1^*\times\vec{\Delta}_1
=-(\vec{\Delta}_1\times\vec{\Delta}_1^*)^*$, $\vec{\Delta}_1\times\vec{\Delta}_1^*$
is pure imaginary. To minimize $\mathrm{Tr}[\Delta\Delta^{\dagger}\Delta \Delta^{\dagger}]$,
we have $\vec{\Delta}_1\times\vec{\Delta}_1^*=\vec{\Delta}_2\times\vec{\Delta}_2^*=0$.
This implies $\vec{\Delta}_1=\vec{d}_1\exp(i\theta_1)$ and $\vec{\Delta}_2=\vec{d}_2\exp(i\theta_2)$
in which $\vec{d}_1$ and $\vec{d}_2$ are real vectors. Then, we obtain
\begin{equation}\label{equa1}
\mathrm{Tr}[\Delta\Delta^{\dagger}\Delta \Delta^{\dagger}]
=\frac{1}{8}(|\vec{d}_1|^4+|\vec{d}_2|^4+\frac{4}{3}|\vec{d}_1|^2|\vec{d}_2|^2)
+\frac{1}{12}\cos[2(\theta_1-\theta_2)]\{(\vec{d}_1\cdot\vec{d}_2)^2-|\vec{d}_1\times\vec{d}_2|^2\}.
\end{equation}
Further minimization requires $\cos[2(\theta_1-\theta_2)]=\pm$1. This constrains
$\theta_1-\theta_2$ while $\theta_1$ could vary freely. In the following we take
$\theta_1=0$ and $\theta_1-\theta_2=\theta$, which leads to $\vec{\Delta}_1=\vec{d}_1$
and $\vec{\Delta}_2=\vec{d}_2\exp(-i\theta)$. When $\cos(2\theta)=1$, we have $\vec{d}_1\perp\vec{d}_2$.
When $\cos(2\theta)=-1$, we have $\vec{d}_1\parallel\vec{d}_2$. In both cases, the
effective action $\mathcal{L^{\prime\prime}}$ becomes:
\begin{small}
\begin{equation}\label{equa1}
\begin{aligned}
\mathcal{L^{\prime\prime}}
&=\{\mathrm{Tr}[G_{+} G_{-}]+\frac{1}{|\lambda_p|}\}(|\vec{d}_1|^2+|\vec{d}_2|^2)
+\frac{\mathrm{Tr}[G_{+}G_{-}G_{+}G_{-}]}{8}(|\vec{d}_1|^4+|\vec{d}_2|^4+\frac{2}{3}|\vec{d}_1|^2|\vec{d}_2|^2) \\
&=\{\mathrm{Tr}[G_{+} G_{-}]+\frac{1}{|\lambda_p|}\}(|\vec{d}_1|^2+|\vec{d}_2|^2)
+\frac{\mathrm{Tr}[G_{+}G_{-}G_{+}G_{-}]}{8}[(|\vec{d}_1|^2+|\vec{d}_2|^2)^2-\frac{4}{3}|\vec{d}_1|^2|\vec{d}_2|^2],
\end{aligned}
\end{equation}
\end{small}
which is minimized when $|\vec{d}_1|^2=|\vec{d}_2|^2$. Note that $\vec{d}_1$ can still
rotate freely.

We now consider the case that $\cos(2\theta)=-1$ and $\vec{d}_1\parallel\vec{d}_2$. 
It is straightforward to obtain
\begin{eqnarray}\label{equa1}
\vec{\Delta}_2
=\pm i\vec{\Delta}_1= (\pm id_{1x},\pm id_{1y},\pm id_{1z}).
\end{eqnarray}
Then the order parameter is:
\begin{eqnarray}\label{equa1}
\Delta
&=&[i(\vec{\Delta}_1\cdot\vec{\sigma})\sigma_y]\otimes[P_x]+[i(\vec{\Delta}_2\cdot\vec{\sigma})\sigma_y]\otimes[P_y]\\
&=&\left(
    \begin{array}{cc}
      -d_{1x}+id_{1y} & d_{1z} \\
      d_{1z} & d_{1x}+id_{1y} \\
    \end{array}
  \right)\otimes[P_x\pm iP_y],
\end{eqnarray}
which corresponds to two chiral $p+ip$ paring channels.

When $\cos(2\theta)=1$, we have $\vec{d}_1\perp\vec{d}_2$ and $\exp(-i\theta)=\pm1$.
Since $\vec{d}_1$ can rotate freely, we take $\vec{d}_{1z}=0$ for simplicity.
$\vec{d}_1\perp\vec{d}_2$ could be fulfilled if we take $d_{2x}=d_{1y}$ and
$d_{2y}=-d_{1x}$. To satisfy $\exp(-i\theta)=\pm 1$, it is clear that $\vec{\Delta}_1=\vec{d}_1$ and
$\vec{\Delta}_2=\pm\vec{d}_2$. Both $\vec{\Delta}_1$ and $\vec{\Delta}_2$ are
real. The order parameter is
\begin{equation}\label{equa1}
\begin{aligned}
\Delta
&=[i(\vec{\Delta}_1\cdot\vec{\sigma})\sigma_y]\otimes[P_x]+[i(\vec{\Delta}_2\cdot\vec{\sigma})\sigma_y]\otimes[P_y] \\
&=\left(
    \begin{array}{cc}
      -d_{1x}+id_{1y} & 0 \\
      0 & d_{1x}+id_{1y} \\
    \end{array}
  \right)\otimes P_x{\pm}\left(
    \begin{array}{cc}
      -d_{1y}-id_{1x} & 0 \\
      0 & d_{1y}-id_{1x} \\
    \end{array}
  \right)\otimes P_y \\
&=\left(
    \begin{array}{cc}
      (-d_{1x}+id_{1y})(P_x{\pm}iP_y) & 0 \\
      0 & (d_{1x}+id_{1y})(P_x{\mp}iP_y) \\
    \end{array}
  \right),
\end{aligned}
\end{equation}
which preserves the time reversal symmetry since both
$\vec{\Delta}_1$ and $\vec{\Delta}_2$ are real. This state corresponds to a helical
$p+ip$ superconductor. The other two helical $p+ip$ paring channels correspond to $d_{2x}=-d_{1y}$ and $d_{2y}=d_{1x}$.

As we have show previously, the effective action $\mathcal{L^{\prime\prime}}$ reaches the same minimum value
\begin{equation}\label{equa1}
\begin{aligned}
\mathcal{L^{\prime\prime}}
&=\{\mathrm{Tr}[G_{+} G_{-}]+\frac{1}{|\lambda_p|}\}(|\vec{d}_1|^2+|\vec{d}_2|^2)
+\frac{\mathrm{Tr}[G_{+}G_{-}G_{+}G_{-}]}{8}(|\vec{d}_1|^4+|\vec{d}_2|^4+\frac{2}{3}|\vec{d}_1|^2|\vec{d}_2|^2) \\
&=\{\mathrm{Tr}[G_{+} G_{-}]+\frac{1}{|\lambda_p|}\}(|\vec{d}_1|^2+|\vec{d}_2|^2)
+\frac{\mathrm{Tr}[G_{+}G_{-}G_{+}G_{-}]}{8}[(|\vec{d}_1|^2+|\vec{d}_2|^2)^2-\frac{4}{3}|\vec{d}_1|^2|\vec{d}_2|^2]
\end{aligned}
\end{equation}
for both cases: (i) $\cos(2\theta)=1$ and $\vec{d}_1\perp\vec{d}_2$ and (ii) $\cos(2\theta)=-1$ and $\vec{d}_1\parallel\vec{d}_2$. This means the two chiral $p+ip$ pairing channels are degenerate with
the four helical $p+ip$ pairing channels, which we have stressed in the main text.
\end{widetext}


\begin{thebibliography}{40}

\bibitem{wenbook} X.-G. Wen, {\it Quantum Field Theory of Many-body Sys-
tems}, Oxford University Press, New York (2004).

\bibitem{hasan-kane-10} M. Z. Hasan, and  C. L. Kane, Rev. Mod. Phys. {\bf 82}, 3045-3067 (2010).

\bibitem{qi-zhang-11} X.-L. Qi, and  S.-C. Zhang, Rev. Mod. Phys. {\bf 83}, 1057-1110 (2011).

\bibitem{qi-hughes-zhang-08} X.-L. Qi, T. L. Hughes, and S.-C. Zhang, Phys. Rev. B {\bf 78}, 195424 (2008).

\bibitem{yuanbo-05}  Y.-B. Zhang, Y. W. Tan, H. L. Stormer, and Kim, Philip. Nature {\bf 438}, 201-204 (2005).

\bibitem{novoselov-05} K. S. Novoselov, A. K. Geim, S. V. Morozov, D. Jiang, M. I. Katsnelson, I. V. Grigorieva, S. V. Dubonos, and A. A. Firsov, Nature {\bf 438}, 197-200 (2005).

\bibitem{qkxue-13} C.-Z. Chang, J. Zhang, X. Feng, J. Shen, Z. Zhang, M. Guo, K. Li, Y. Ou, P. Wei, L.-L. Wang, Z.-Q. Ji, Y. Feng, S. Ji, X. Chen, J. Jia, X. Dai, Z. Fang, S.-C. Zhang, K. He, Y. Wang, L. Lu, X.-C. Ma, Q.-K. Xue, Science {\bf 340}, 167-170 (2013).

\bibitem{kitaev-09} A. Kitaev, AIP Conf. Proc. {\bf 1134}, 22-30 (2009).

\bibitem{schnyder-08} A. P. Schnyder, S. Ryu, A. Furusaki, and  A. W. W. Ludwig,Phys. Rev. B {\bf 78}, 195125 (2008).

\bibitem{Qi2}  X.-L. Qi, T. L. Hughes, S. Raghu, and S.-C. Zhang, Phys. Rev. Lett. {\bf102}, 187001 (2009).

\bibitem{Sigrist-95}  T. M. Rice, and M. J. Sigrist, Phys. Condens. Matter {\bf 7}, L643-L648 (1995).

\bibitem{Mackenzie}  A. P. Mackenzie, and Y. Maeno, Rev. Mod. Phys. {\bf75}, 657-712 (2003).

\bibitem{sri-kapitulnik-kivelson} S. Raghu, A. Kapitulnik, and S. A. Kivelson, Phys. Rev. Lett. {\bf 105}, 136401 (2010).

\bibitem{ronny-13} Q. H. Wang, C. Platt, Y. Yang, C. Honerkamp, F. C. Zhang, W. Hanke, T. M. Rice, and R. Thomale, Europhys. Lett. {\bf 104}, 17013 (2013).


\bibitem{Tewari-Sarma-Lee-08} S. Tewari, S. Das Sarma, and D. H. Lee, Phys. Rev. Lett. {\bf 99},037001 (2007).

\bibitem{Read-Green}  N. Read, and D. Green, Phys. Rev. B {\bf61}, 10267-10297 (2000).

\bibitem{Ivanov} D. A. Ivanov, Phys. Rev. Lett. {\bf86}, 268-271 (2001).

\bibitem{kitaev-03} A. Y. Kitaev, Ann. Phys. {\bf 303}, 2-30 (2003).

\bibitem{Nayak}  C. Nayak, S. H. Simon, A. Stern, M. Freedman, and S. Das Sarma, Rev. Mod. Phys. {\bf80}, 1083-1159 (2008).

\bibitem{review-12} C. Kallin, Rep. Prog. Phys. {\bf 75}, 042501 (2012).




\bibitem{novoselov-04}  K. S. Novoselov, {\it et al.}. Science {\bf 306}, 666-669 (2004).

\bibitem{castroneto-09} A. H. Castro Neto, F. Guinea, N. M. R. Peres, K. S. Novoselov, and A. K. Geim, Rev. Mod. Phys. {\bf 81}, 109-162 (2009).


\bibitem{uchoa-07} B. Uchoa, and A. H. Castro Neto, Phys. Rev. Lett. {\bf 98}, 146801 (2007).

\bibitem{sri-kivelson-scalapino}  S. Raghu, S. A. Kivelson, and D. Scalapino, J. Phys. Rev. B {\bf 81}, 224505 (2010).

\bibitem{chubukov-12} R. Nandkishore, L. S. Levitov, and A. V. Chubukov, Nat. Phys. {\bf 8}, 158-163 (2012).

\bibitem{dhlee-121} W.-S. Wang, Y.-Y. Xiang, Q.-H. Wang, F. Wang, F. Yang, and D.-H. Lee, Phys. Rev. B {\bf 85}, 035414 (2012).

\bibitem{ronny-12} M. L. Kiesel, C. Platt, W. Hanke, D. A. Abanin, and R. Thomale, Phys. Rev. B {\bf 86}, 020507(R) (2012).

\bibitem{yao-13} H. Yao, and F. Yang, Phys. Rev. B {\bf 92}, 035132 (2015).

\bibitem{Yanagisawa}  H. Yanagisawa, T. Tanaka, Y. Ishida, M. Matsue, E. Rokuta, S. Otani, and C. Oshima, Phys. Rev. Lett. {\bf93} , 177003 (2004).

\bibitem{Kresse1}   G. Kresse, and J. Furthm\"{u}ller, Phys. Rev. B {\bf54}, 11169-11186 (1996).

\bibitem{Kresse2}   G. Kresse, and J. Furthm\"{u}ller, Comput. Mater. Sci. {\bf 6}, 15-50 (1996).

\bibitem{Miyamoto}  Y. Miyamoto, A. Rubio, S. G. Louie, and M. L. Cohen, Phys. Rev. B {\bf50}, 18360 (1994).

\bibitem{Chen}    X. Chen, and J. Ni, Phys. Rev. B {\bf88}, 115430 (2013).

\bibitem{Scalapino89} N. E. Bickers, D. J. Scalapino, and S. R. White, Phys. Rev. Lett. {\bf 62}, 961 (1989).




\bibitem{RMP} W. Metzner, M. Salmhofer, C. Honerkamp, V. Meden, and K. Schonhammer, Rev. Mod. Phys.{\bf84}, 299(2012).

\bibitem{Furukawa}   N. Furukawa, T. M. Rice, and M. Salmhofer, Phys. Rev. Lett. {\bf81}, 3195-3198 (1998).

\bibitem{liu-2013}  F. Liu, C.-C. Liu, K. H. Wu, F. Yang, and Y. G. Yao, Phys. Rev. Lett. {\bf111}, 066804 (2013).

\bibitem{Zhang} J. Zhang, C. Lorscher, Q. Gu and R. A. Klemm, J. Phys.: Condens. Matter 26, 252201 (2014).

\bibitem{SQUID} D. J. Van Harlingen, Rev. Mod. Phys. \textbf{67}, 515--535 (1995).


\bibitem{dhlee-122} Y.-Y. Xiang, W.-S. Wang, Q.-H. Wang, and D.-H. Lee, Phys. Rev. B {\bf86}, 024523 (2012).

\bibitem{Nagaosa}  S. Nakosai, Y. Tanaka, and N. Nagaosa, Phys. Rev. Lett. {\bf108}, 147003 (2012).

\bibitem{shoucheng-14} J. Wang, Y. Xu, and S.-C. Zhang, Phys. Rev. B {\bf90}, 054503 (2014).

\bibitem{yang-14} F. Yang, C.-C. Liu, Y.-Z. Zhang, Y. Yao, and D.-H. Lee, Phys. Rev. B {\bf91}, 134514 (2015).

\bibitem{Grover}  T. Grover, D. N. Sheng,and A. Vishwanath, Science {\bf 344}, 280 (2014).

\bibitem{sslee} P. Ponte and S.-S. Lee, New J. Phys. {\bf 16}, 013044 (2014).

\bibitem{yao-14} S.-K. Jian, Y.-F. Jiang, and H. Yao, Phys. Rev. Lett. {\bf 114}, 237001 (2015).


\end{thebibliography}
\end{document}